\documentclass[doublecol,linenumbers]{epl2}
\usepackage{graphicx}
\usepackage[english]{babel}
\usepackage{latexsym,amsmath,amscd,amssymb,graphicx,amsbsy,color,xcolor,amsfonts,amsthm}
\title{Relativity from the Geometrization of Newtonian Dynamics}
\author{Yaakov Friedman and Tzvi Scarr}
\institute{Jerusalem College of Technology\\Jerusalem, Israel}


\abstract{ Based on the \emph{Generalized Principle of Inertia}, which states that: \emph{An inanimate object moves freely, that is, with zero acceleration, in its own spacetime, whose geometry is determined by 
all of the forces affecting it,} we geometrize Newtonian dynamics for any conservative force. For an object moving in a spherically symmetric
  force field, using a variational principle, conservation of angular momentum and a classical limit, we construct a metric with respect to which the object's worldline is a geodesic.  For the 
  gravitational field of a static, spherically symmetric mass, this metric is the Schwarzschild metric. The resulting dynamics reduces in the weak field, low velocity limit to classical Newtonian dynamics 
  and exactly reproduces the classical tests of General Relativity. The metric of gravitoelectromagnetism is extended to handle a gravitational  field generated by several sources.}

\begin{document}
\maketitle

\section{Introduction}\label{intro}
$\;$
Bernhard Riemann was a strong advocate of a geometric approach to physics. As pointed out in \cite{papa}, ``one of the main features of the local geometry conceived by Riemann is that it is well suited to the study of gravity and more general fields in physics." He believed that the forces at play in a system determine the geometry of the system. For Riemann, \emph{force equals geometry}.

General Relativity ($GR$) is a direct application of ``force equals geometry." In $GR$, the gravitational force curves spacetime. Since, by the Equivalence Principle, the acceleration of an object in a gravitational field is independent of its mass, curved spacetime can be considered a \emph{stage} on which objects move. In other words, the geometry is the same for all objects. However, the Equivalence Principle holds only for gravitation. In this way, $GR$ singles out the gravitational force from other forces which are not treated geometrically. For example, the potential of an electric force depends on the charge of the particle, and the particle's acceleration depends on its charge-to-mass ratio. Thus, the electric field does not create a common stage on which all particles move. Indeed, a neutral particle does not feel any electric force at all. The way spacetime curves due to an electric field depends on \emph{both} the field \emph{and} intrinsic properties of the object. This was also recognized in the geometric approach of \cite{Duarte}. How, then, are we to apply Riemann's principle of ``force equals geometry" to other forces?

We answer this question here for any conservative force. One of the main new ideas is the \emph{relativity of spacetime}. This means that spacetime is an object-dependent notion. An object lives in its own spacetime, its own geometric world, which is defined by the forces which affect it. For example, in the vicinity of an electric field, a charged particle and a neutral particle exist in different worlds, in different spacetimes. In fact, for the neutral particle, the electric field does not exist. Likewise, in the vicinity of a magnet, a piece of iron and a piece of plastic live in two different worlds.

An inanimate object has no internal mechanism with which to change its velocity. Hence, it has constant velocity, or zero acceleration, in \emph{its own} world (spacetime). This leads us to formulate a new principle, the \emph{Generalized Principle of Inertia}, which unifies Newton's first and second laws and states that:  \textbf{An inanimate object moves freely, that is, with zero acceleration, in \emph{its own} spacetime, whose geometry is determined by all of the forces affecting it}. This is a generalization, or more accurately, a relativization of  Einstein's idea.

In $GR$, a freely-falling object  in a gravitational field moves along a geodesic determined by the metric of the spacetime. With respect to this metric, the object's acceleration is zero. The Generalized Principle of Inertia extends this idea so that \emph{every} object moves along a geodesic in \emph{its} spacetime.  This geodesic is with respect to an appropriate metric, which we call the \emph{metric of the object's spacetime}. The metric in $GR$ is determined solely by the gravitational sources, while in our model, the metric of the object's spacetime depends on all of the forces affecting the object. In the case of static, conservative forces, the metric will depend only on the potentials of these forces, defined in some preferred inertial frame, which may be the rest frame of the Universe.

%This suggests that the motion of an object under the given force may be represented by geodesics.  We assume that the object's spacetime has a metric with respect to which the length of the object's trajectory is extremal.   We call this metric the

Since the motion is by geodesics, we use a variational principle and derive Euler-Lagrange type equations and the ensuing conservation laws.
\begin{quotation}
``Many results in both classical and quantum physics can be expressed as variational principles, and it is often when expressed in this form that their physical
meaning is most clearly understood. Moreover, once a physical phenomenon has been written as a variational principle, ... it is usually possible to identify conserved
quantities, or symmetries of the system of interest, that otherwise might be found only with considerable effort \cite{Riley}."
\end{quotation}
In the classification of alternative gravitation theories of \cite{Will1}, our theory is a preferred frame, Lagrangian-based metric theory.

The metric for static, conservative forces obtained here is the same as that obtained in \cite{FRND,FStProc} using escape trajectories to describe the influence of the field on spacetime. This metric is called the \emph{Relativistic Newtonian Dynamics (RND)} metric. It is also the same as the metric derived in \cite{FSS}. There, the authors first derive a Newtonian metric and then correct it in order to handle relativistic effects. In this paper, we obtain the \emph{RND} metric directly from symmetries and the Newtonian limit.

We start by restricting our attention to the motion of an object in a static, spherically symmetric, attractive central force field. The explicit form of the metric of the object's spacetime is derived from the above Euler-Lagrange equations and  conservation laws, in addition to spherical symmetry and a carefully defined classical limit. For the gravitational force, this metric turns out to be the Schwarzschild metric. We thus obtain a simple derivation of the Schwarzschild metric, one that does not require the field equations of $GR$.

This metric is generalized to any static, conservative force and a formula for the total energy is obtained.  The energy equation contains the usual kinetic and potential energy terms but also a new term which includes both kinetic and potential energy. This implies that in order to reproduce relativistic effects, one can no longer separate these contributions. These results  are applied to the case of the gravitational field of a static, spherically symmetric mass distribution and exactly reproduce the classical tests of $GR$. Finally, we also consider the non-static case and obtain a candidate for the metric of the spacetime of an object in a field generated by several sources.

\section{A geometric approach to dynamics}\label{opti}
$\;$
We present the preferred frame, Lagrangian approach to geometrize the motion of an object of mass $m$ in a force field. The results here are quite general and may be applied to any force field. Later, we will consider the particular case of a static, spherically symmetric, attractive force field.

Let
\begin{equation}\label{arbmet}
ds^2=g_{\alpha\beta}(q)dq^\alpha dq^\beta
\end{equation}
be the metric of the object's spacetime, where $q^\alpha, \alpha=0,1,2,3$ are the coordinates of an inertial observer far removed from the sources of the field. This is similar to Earth-based observations of motion within a distant galaxy and to observations within the solar system measured with respect to the far-removed stars.

In order to define the length of the trajectory, we introduce the following function $L$ of the eight independent variables $q^\alpha,\acute{q}^\alpha$:
\begin{equation}\label{L}
L\left(q,\acute{q}\right)=mc\sqrt{g_{\alpha\beta}(q)\acute{q}^\alpha\acute{q}^\beta},
\end{equation}
where the constant $mc$ has been chosen to be compatible with the classical notion of momentum.

 Let $q:\sigma \to x, a\le\sigma\le b$ be a trajectory of an object of mass $m$, where $\sigma$ is an arbitrary parameter. The length $l(q)$ of the trajectory $q$ is given by
\begin{equation}\label{lq}
l(q)=mc\int_a^b \frac{ds}{d\sigma}d\sigma=\int_a^b L\left(q,\frac{dq}{d\sigma}\right)d\sigma.
\end{equation}
It is well known that the length of the trajectory does not depend on the parametrization.
%Let $u:\sigma \to x, a\le\sigma\le b, u(a)=u(b)=0$ be a perturbation of the trajectory. The length of $q(\sigma)$ is extremal if
%\begin{equation}\label{extcon}
%\frac{d}{d\epsilon}l(q+\epsilon u)|_{\epsilon=0}=0.
%\end{equation}

From the Generalized Principle of Inertia, the length of the trajectory $q(\sigma)$ is extremal. By a standard argument, it follows that $q(\sigma)$ satisfies the Euler-Lagrange equations
\begin{equation}\label{EL eqns}
\frac{\partial L}{\partial q^\mu}-\frac{d}{d\sigma}\frac{\partial L}{\partial \acute{q}^\mu}=0,
\end{equation}
where $L$ is defined by (\ref{L}) and $\acute{q}=\frac{dq}{d\sigma}$. In this case, the conjugate momentum $p_\mu$ is
\begin{equation}\label{compMoment}
  p_\mu = \frac{\partial L}{\partial\acute{q}^\mu}\Bigr|_{\acute{q}=\frac{dq}{d\sigma}}  =\frac{mcg_{\mu\beta}\frac{dq^\beta}{d\sigma}}{ds/d\sigma}=mcg_{\mu\beta}\frac{dq^\beta}{ds}.
\end{equation}

Note that the second term in equation (\ref{EL eqns}) contains differentiation by two parameters on the curve. The first differentiation is by $s$, as seen in equation (\ref{compMoment}). The second differentiation is by $\sigma$. In order to
obtain a differential equation with a single parameter, we will choose $\sigma$ to be proportional to $s$. More precisely, we choose $\sigma$ to be the parameter
\begin{equation}\label{tauis}
\tau=c^{-1}s,
\end{equation}
called \emph{proper time}, which is proportional to $s$ and reduces to the coordinate time $t$ in the classical limit.

Using $\tau$ will turn equation (\ref{EL eqns}) into a second-order differential equation. We denote differentiation of $q$ with respect to $\tau$ by $\dot{q}$.
From (\ref{EL eqns}) and (\ref{compMoment}), the Euler-Lagrange equations become
\begin{equation}\label{EOM}
\frac{\partial L}{\partial q^\mu}=\frac{dp_\mu}{d\tau},
\end{equation}
where
\begin{equation}\label{pitau}
p_\mu=mg_{\mu\beta}\dot{q}^\beta.
\end{equation}
Note that momentum is a covector, while velocity is a vector. Therefore, the metric is necessary in (\ref{pitau}) in order to lower the index. It can be shown that (\ref{EOM}) is equivalent to the geodesic equation.

The following proposition follows immediately from equation (\ref{EOM}).
\vskip0.3cm
\noindent\textbf{Proposition 1}\;\;\; If the metric coefficients $g_{\alpha\beta}$ do not depend on the coordinate $q^\mu$, then the $\mu$ component $p_\mu=mg_{\mu\beta}\dot{q}^\beta$ of the conjugate momentum is conserved on the
trajectory.

\section{The metric of an object moving in a static, spherically symmetric, attractive central force field}\label{staticss}
$\;\;\;$
Now we derive the metric of the spacetime of an object of mass $m$ moving in a static, spherically symmetric, attractive central force field.  Let $K$ be of an inertial observer far removed from the sources of the field. We use standard
spherical coordinates $ct,r,\theta,\varphi$   with the origin at  the center of symmetry of the field in $K$.  Let $U(r)$ denote the potential of this field. We assume that $U\leq 0$ and vanishes at infinity.

\subsection{The implications of spherical symmetry}

We turn now to the construction of the metric of the object's spacetime. From spherical symmetry, the metric in $K$ is of the form
\begin{equation}\label{genmet}
ds^2=f(r)c^2dt^2-g(r)dr^2-h(r)r^2(d\theta^2+\sin^2\theta d\varphi^2).
\end{equation}
 Since the force is static, the metric coefficients do not depend on $t$. Moreover, there are no time-space cross terms (see \cite{Rindler}, page 187). By spherical symmetry, the functions $f,g$ and $h$ cannot depend on
$\varphi$. Since $U(r)$ vanishes at infinity,  the metric is asymptotically the Minkowski metric
\begin{equation}\label{MinkMet}
\eta_{\alpha\beta}=\operatorname{diag}(1,-1,-1,-1).
\end{equation}
Hence,
\begin{equation}\label{normal}
 \lim_{r\rightarrow\infty}f(r)=\lim_{r\rightarrow\infty}g(r)=\lim_{r\rightarrow\infty}h(r)=1.
\end{equation}

Without loss of generality, we may assume that the motion is in the plane $\theta=\pi/2$. Since the metric coefficients do not depend on $\varphi$, Proposition 1 implies that the $\varphi$ component of the conjugate momentum is conserved on the trajectory of the object. Thus,
\begin{equation}\label{Kcond2IND2}
h(r)r^2\dot{\varphi}=J,
\end{equation}
for some constant $J$. Since the force is central, the acceleration $\ddot{\mathbf{r}}$ is parallel to $\mathbf{r}$ implying that the angular momentum per unit mass $\mathbf{r}\times\dot{\mathbf{r}}=r^2\dot{\varphi}$  is also conserved on the trajectory. Thus $h(r)$ remains constant for any values $r$ on the trajectory of the object. Since around any value $r>0$ there can be a trajectory of the object, this implies that $h(r)$ is constant for all $r$, and from (\ref{normal}),
\begin{equation}\label{his1}
h(r)=1.
\end{equation}

\subsection{Newtonian approximation for radial motion}

Since angular momentum conservation does not provide any information about radial motion ($\varphi=$ constant), we now consider radial motion.
The function $L(t,r,\dot{t},\dot{r})$ defined by (\ref{L}) in this case becomes
\begin{equation}\label{Lr}
L\left(t,r,\dot{t},\dot{r}\right)=mc\sqrt{f(r)c^2\dot{t}^2-g(r)\dot{r}^2}.
\end{equation}
From (\ref{pitau}), the $r$-momentum is $p_r=-mg(r)\dot{r}$, and the $r$ component of the  Euler-Lagrange equation (\ref{EOM}) is
\begin{equation}\label{ELRad}
\frac{m}{2}\left(f'(r)c^2\dot{t}^2-g'(r)\dot{r}^2 \right)+mg'(r)\dot{r}^2+mg\ddot{r} =0.
\end{equation}
Dividing (\ref{genmet}) by $d\tau^2$, we obtain the norm equation
\begin{equation}\label{normeqnrad}
f(r)c^2\dot{t}^2-g(r)\dot{r}^2=c^2,
\end{equation}
implying that $c^2\dot{t}^2=\frac{c^2}{f(r)} +\frac{g(r)}{f(r)}\dot{r}^2$. Substituting this into (\ref{ELRad}), we obtain
\begin{equation}\label{RadEq}
\ddot{r} =-\frac{1}{2}\left(\left(\frac{f'(r)}{f(r)}+\frac{g'(r)}{g(r)} \right)\dot{r}^2+\frac{c^2f'(r)}{f(r)g(r)}\right).
\end{equation}

In Newtonian dynamics, the acceleration of an object depends only on the forces acting on it and on the object's mass, but is independent of its velocity. As explained above, in the geometric approach to dynamics, the acceleration should be $\ddot{r}=\frac{d^2r}{d\tau^2}$. Can we define the metric so that this acceleration for radial motion will be independent on the velocity of the object? From equation (\ref{RadEq}), it follows that this can be achieved if we set
\begin{equation}\label{choise}
\frac{f'(r)}{f(r)}+\frac{g'(r)}{g(r)}=0,
\end{equation}
which implies that $\ln (f(r)g(r))=const$ and $f(r)g(r)=const$. Using (\ref{normal}), this implies that
\begin{equation}\label{Tan}
  f(r)g(r)=1.
\end{equation}
The condition (\ref{Tan}) was also used in \cite{Tang}. For some interesting conditions equivalent to (\ref{Tan}), see \cite{GuenGRG} and \cite{Jacob}. Under this condition, (\ref{RadEq}) becomes
\begin{equation}\label{RadEq2}
\ddot{r} =-\frac{c^2f'(r)}{2}.
\end{equation}

The function $f(r)$ will be defined from the classical limit. Let $r_0$ be an arbitrary value of $r$. Consider the radial motion of an object whose velocity at $r_0$ is $v(r_0)=0$. We now connect this object's acceleration $\ddot{r}$ with the classical acceleration $\frac{d^2r}{dt^2}$ at $r_0$. Since near the point $r_0$, we have $d\tau=\sqrt{f(r)}dt$, the acceleration $\ddot{r}$ at $r_0$ is
\begin{equation*}\label{ddot123}
 \ddot{r}=\frac{d^2r}{d\tau^2}=\frac{d}{d\tau}\left(\frac{1}{\sqrt{f}}\frac{dr}{dt}\right)
\end{equation*}
\begin{equation}\label{ddot124}
=\left(-\frac{1}{2}f^{-3/2}f'\left(\frac{dr}{dt}\right)^2 +\frac{1}{\sqrt{f}}\frac{d^2r}{dt^2}\right)f^{-1/2}.
\end{equation}
Since $v(r_0)=0$, we have
\begin{equation}\label{accr0}
\ddot{r}(r_0)=\frac{1}{f(r_0)}\frac{d^2r}{dt^2}\Bigr|_{r_0}.
\end{equation}

The Newtonian radial acceleration in tensorial form is $\frac{d^2r}{dt^2}=m^{-1}\eta^{1\beta}U,_\beta$ (see \cite{Itin}). Since the object's spacetime is not flat, this formula should be replaced by $\frac{d^2r}{dt^2}=m^{-1}g^{1\beta}U,_\beta$. For radial motion with metric (\ref{genmet}) satisfying (\ref{Tan}), this acceleration is $\frac{d^2r}{dt^2}=-\frac{1}{mg(r)}\frac{dU}{dr}=-\frac{f(r)}{m}\frac{dU}{dr}$. Hence, the acceleration $\ddot{r}$ at $r_0$ is
\begin{equation}\label{Geom2New}
\ddot{r}(r_0)=\frac{1}{f(r_0)}\frac{d^2r}{dt^2}\Bigr|_{r_0}=-\frac{1}{f(r_0)}\frac{f(r_0)}{m}\frac{dU}{dr}\Bigr|_{r_0}=-\frac{1}{m}\frac{dU}{dr}\Bigr|_{r_0}.
\end{equation}
This is the analog of Newton's second law which, for geometric dynamics, holds only in the case in which the velocity is in the direction of the acceleration.

 Equation (\ref{Geom2New}) holds for any radial motion, not just motion with has zero velocity at some point, since the acceleration $\ddot{r} $ which satisfies (\ref{RadEq2}) is independent of the velocity of the object. Moreover, since $r_0$ was arbitrary, the radial acceleration is
\begin{equation}\label{Geom2New2}
\ddot{r}=-\frac{1}{m}\frac{dU}{dr}.
\end{equation}
For radial motion, the geometrization of Newton's second law involves only replacing the time parameter with the proper time parameter.

From (\ref{RadEq2}), we obtain $\frac{c^2f'(r)}{2}=\frac{\frac{dU}{dr}}{m}$. Thus,
$ f(r)=\frac{2U}{mc^2}+const$, and from (\ref{normal}), $f(r)=1+\frac{2U}{mc^2}$.  Introducing the \textit{dimensionless potential}
\begin{equation}\label{u_def}
u(r)=\frac{-2U(r)}{mc^2},
\end{equation}
we have
\begin{equation}\label{fDef}
f(r)=1-u(r),
\end{equation}
and from (\ref{Tan}),
\begin{equation}\label{gDef}
 g(r)=\frac{1}{1-u(r)}.
\end{equation}

Thus,  using $u(r)$ defined by (\ref{u_def}), the metric (\ref{genmet}) is
\begin{equation}\label{fullmetric}
ds^2=(1-u(r))c^2dt^2-\frac{1}{1-u(r)}dr^2- r^2(d\theta^2+\sin^2\theta d\varphi^2),
\end{equation}
which in the case of the gravitational field of a spherically symmetric mass distribution is the Schwarzschild metric. This metric was also derived without using Einstein's equation in \cite{FRND,FStProc} and also in \cite{GuenBulg}, in each case, from assumptions different from those used here.

Our approach can be generalized to motion under any conservative force with potential $ U(\mathbf{x})$, as follows.  Introduce at each $\mathbf{x}$ where $\nabla U(\mathbf{x})\neq \mathbf{0}$  a normalized vector
  \begin{equation}\label{n-def}
 \mathbf{n}(\mathbf{x}) = \frac{\nabla U(\mathbf{x})}{|\nabla U(\mathbf{x})|}
\end{equation}
 in the direction of the gradient of $U(\mathbf{x})$, or the negative of the direction of the force. Let $d\mathbf{x}_n=(d\mathbf{x}\cdot\mathbf{n})\mathbf{n}$ and  $d\mathbf{x}_{tr}=d\mathbf{x}-(d\mathbf{x}\cdot\mathbf{n})\mathbf{n}$,  respectively, denote the projections of the spatial increment $d\mathbf{x}$ in the parallel and transverse directions to $\mathbf{n}(\mathbf{x})$. With this notation, our metric (\ref{fullmetric}) becomes
 \begin{equation}\label{normRND}
 ds^2=(1-u(\mathbf{x}))c^2dt^2-\frac{1}{1-u(\mathbf{x})} d\mathbf{x}_n^2-d\mathbf{x}_{tr}^2,
\end{equation}
which is the $RND$ metric.

\section{Energy and Evolution Equations of $RND$ in a conservative force field}\label{energy}
$\;$

%The above derivation of the metric could be extended in a similar way for motion under any attractive static conservative force with potential $U(\mathbf{x})$ for $\mathbf{x}\in R^3$, vanishing at infinity. Denote by $\mathbf{n}$ is a unit vector in the direction of $\nabla U$ and decompose $d\mathbf{x}$ into two components $d\mathbf{x}_n=(d\mathbf{x}\cdot\mathbf{n})\mathbf{n}$ and $d\mathbf{x}_{tr}=d\mathbf{x}-d\mathbf{x}_n$. In this case, the metric  (\ref{fullmetric}) is
%\begin{equation}\label{fullmetric2}
%ds^2=(1-u(\mathbf{x}))c^2dt^2-\frac{1}{1-u(\mathbf{x})}d\mathbf{x}_n^2- d\mathbf{x}_{tr}^2,
%\end{equation}
%where, as in (\ref{u_def}),  $u(\mathbf{x})=-\frac{2U(\mathbf{x})}{mc^2}.$

We now obtain the dimensionless and dimensional \textit{energy conservation equations} and \textit{equations of motion} of $RND$. Since the metric (\ref{normRND}) is static, Proposition 1 implies  that
\begin{equation}\label{Kcond3IND}
\dot{t}=\frac{k}{1-u(\mathbf{x})}
\end{equation}
for some constant $k$. The square of the norm of the four-velocity $\dot{x}=(c\dot{t},\dot{\mathbf{x}})$ with respect to (\ref{normRND}) is
\begin{equation}\label{RNDnormEqn}
c^2\frac{k^2}{1-u}-\frac{\dot{\mathbf{x}}_n^2}{1-u}-\dot{\mathbf{x}}_{tr}^2=c^2,
\end{equation}
where $\dot{\mathbf{x}}_n$ and $\dot{\mathbf{x}}_{tr}$ are, respectively, the radial and transverse components of the velocity. Multiplying by $\frac{1-u}{c^2}$ and rearranging terms, we obtain the  \textit{dimensionless energy conservation equation}
\begin{equation}\label{GNGdimlessenercons}
\frac{\dot{\mathbf{x}}_n^2+(1-u)\dot{\mathbf{x}}_{tr}^2}{c^2}-u=k^2-1.
\end{equation}

Multiplying the previous equation by $\frac{mc^2}{2}$ and using $\dot{\mathbf{x}}^2=\dot{\mathbf{x}}_n^2+\dot{\mathbf{x}}_{tr}^2$, we obtain the corresponding  \textit{dimensional energy conservation equation}
\begin{equation}\label{GNGdimenercons}
   \frac{m\dot{\mathbf{x}}^2}{2}+U(r)\frac{\dot{\mathbf{x}}_{tr}^2}{c^2}+U(\mathbf{x})=E,
\end{equation}
where the integral of motion $E$ is the total energy on the worldline. For radial motion this is similar to the Newtonian energy conservation, but in general, in addition to the usual kinetic and potential energy terms, equation (\ref{GNGdimenercons}) has a \emph{mixed term} which depends on both the velocity of the object and the potential. This means that in order to reproduce relativistic effects, one can no longer distinguish between potential and kinetic energy, as in Newtonian dynamics. This also explains the need to include the velocity in the modified Newtonian
potentials proposed in \cite{TR,Ghosh14,SG,G15,Ghosh16}. The mixed term in (\ref{GNGdimenercons}) is approximately $\beta^2U(\mathbf{x})$ and is therefore only seen for high velocities or in high-precision experiments.

Note that we can also write (\ref{GNGdimenercons}) as
\begin{equation}\label{GNGdimenercons2}
\frac{m}{2}\left(\dot{\mathbf{x}}_n^2+(1-u)\dot{\mathbf{x}}_{tr}^2\right)+U(\mathbf{x})=E.
\end{equation}
This is the usual ``kinetic plus potential" energy from of energy conservation, only now the square of the velocity is computed with respect to the metric (\ref{normRND}). The kinetic energy is now dependent on the both the
position and the velocity of the object.

Differentiating (\ref{GNGdimenercons}) by $\tau$, we obtain the $RND$ evolution equation for a conservative force field:
\begin{equation}\label{RND1finals}
 m\ddot{\mathbf{x}}=-\nabla U-\nabla U\frac{\dot{\mathbf{x}}^2_{tr}}{c^2}+ 2\frac{U(\mathbf{x})}{c^2}(\dot{\mathbf{x}}\cdot\dot{\mathbf{n}})\mathbf{n},
\end{equation}
where $\mathbf{n}$ is a unit vector in the direction of $\nabla U$. This equation has two additional terms not appearing in the corresponding classical equation. In the classical regime, both of these terms are small and have therefore gone
unrecognized.
Details of the derivation of (\ref{RND1finals}) can be found in \cite{FRND} and \cite{FStProc}.

\section{Motion in the Gravitational Field of a spherically symmetric mass distribution}\label{Mercury}
$\;$
We now derive the evolution equation of an object of mass $m$ moving in the gravitational field of a static, spherically symmetric mass distribution $M$. Without loss of generality, we may assume that the motion is in the plane
$\theta=\pi/2$.  Let $U=U(\mathbf{r})=-\frac{GMm}{r}$ denote the potential of this field. Note that $U\leq 0$ and vanishes at infinity. Here,
\begin{equation}\label{ugrav}
u(r)=\frac{-2U(r)}{mc^2}=\frac{r_s}{r},
\end{equation}
where
\begin{equation}\label{Schwarz}
r_s=2GM/c^2
\end{equation}
is the \textit{Schwarzschild radius} .

From (\ref{Kcond2IND2}) and (\ref{his1}), we have $r^2\dot{\varphi}=J$.  Rewrite (\ref{RNDnormEqn}) as
\begin{equation}\label{subIOM}
c^2\frac{k^2}{1-u}-\frac{\dot{r}^2}{1-u}-\frac{J^2}{r^2}=c^2.
\end{equation}
Multiply by $\frac{1-u}{c^2}$ and rearranging terms, we get
\begin{equation}\label{energyEQN}
\frac{\dot{r}^2}{c^2}+(1-u)\left(\frac{J^2}{c^2r^2}+1\right)=k^2.
\end{equation}

We now change variables from $r$ and $\tau$ to $u$ and $\varphi$. We do this in order to describe the trajectory as $u(\varphi)$ or $r(\varphi)$, independently of $\tau$. Let $u'=\frac{du}{d\varphi}$. Using $u=r_s/r$, we have
$\frac{du}{d\tau}=\frac{-r_s}{r^2}\dot{r}$. Using this and $\frac{d\varphi}{d\tau}=\frac{J}{r^2}$, we have
\begin{equation}\label{RNDdrdt}
u'=\frac{du}{d\tau}\frac{d\tau}{d\varphi}=\frac{-r_s}{r^2}\dot{r}\frac{r^2}{J}=\frac{-r_s}{J}\dot{r}\quad,\quad \dot{r}=\frac{-J}{r_s}u'.
\end{equation}
Thus, equation (\ref{energyEQN}) becomes
\begin{equation}\label{energyEQN2}
\frac{J^2}{c^2r_s^2}(u')^2+(1-u)\left(\frac{J^2u^2}{c^2r_s^2}+1\right)=k^2.
\end{equation}
Let
\begin{equation}\label{defmu}
\mu=\frac{r_s^2c^2}{2J^2}.
\end{equation}
Multiply equation (\ref{energyEQN2}) by $2\mu$ and rearrange terms to obtain
\begin{equation}\label{RNDeqnmu}
(u')^2=u^3-u^2+2\mu u+2\mu(k^2-1).
\end{equation}
This equation differs from the corresponding classical equation only in the appearance of the term $u^3$. Since $u\ll 1$, the solution of this equation is a perturbation of the classical solution. Nevertheless, from (\ref{RNDeqnmu}), one can obtain all of the tests of $GR$, assuming that the Sun is at rest in our inertial frame $K$ (see \cite{FSMerc} ,\cite{FSBin}, \cite{FRND}  and \cite{FSLens} ).

\section{The $RND$ metric of several objects moving  in an inertial frame}\label{many}

If the inertial frame $K$ is the rest frame of the Universe, and the rest frame $K'$ of the Sun moves uniformly with respect to $K$, then as pointed out in \cite{Will2}, the theory will pass the tests of $GR$ only if it is covariant
between inertial frames. Therefore, we now extend the previous results by imposing Lorentz covariance. In electromagnetism, the potential of a moving charge can be defined from the potential of a stationary charge by applying Lorentz
covariance. This leads to the Li\'{e}nard-Wiechert  retarded four-potential. Our approach here is similar.

Consider a spherically symmetric object of rest mass $M$ moving uniformly in an inertial frame $K$.

% positioned at the origin at time $t=0$. Denote by $K'$ the inertial frame comoving with the object. Since in $K'$ the object is at rest at the origin, the Newtonian potential at a spacetime point $x'$ is $U'(x')=GM/r'=r_s/r'$, were $r'$ is the spacial distance  of $x'$ from the origin. Since a gravitational field propagate with the speed of light, we may assume that $x'$ a null-vector and thus, $r'$ is equal to the zero component of $x'$. In $K'$ the object is at rest implying that its four-velocity is $w'=(1,0,0,0)$ and  that $r'=x'\cdot w'$, where $\cdot$ denote the Minkowski dot product.  In a $4D$ spacetime the potential should be $4D$ co-vector. Thus, we can write the  four-potential of the gravitational field in $K'$  as
%\begin{equation}\label{4potInrest}
%  A'_\alpha(x')=\frac{-MGw'_\alpha}{x'\cdot w'}.
%\end{equation}
%
%Assume now that the four potential  should be Lorentz covariant.  Denote by $\Lambda$ the Lorentz transformation from $K'$ to $K$. In $K$, the spacetime point  $x'$ become the 4-vector $x=\Lambda x'$ describing the relative position of the observer with respect to the object at the retarded time and $w=\Lambda w'$ is its four-velocity at this time. Using that the dot product is Lorentz invariant, the formula for the four-potential (\ref{4potInrest})  in $K$  is
%\begin{equation}\label{4pot}
%  A_\alpha(x)=\frac{-MG w_\alpha}{x\cdot w}.
%\end{equation}

\begin{figure}[h!]
  \centering
\scalebox{0.27}{\includegraphics{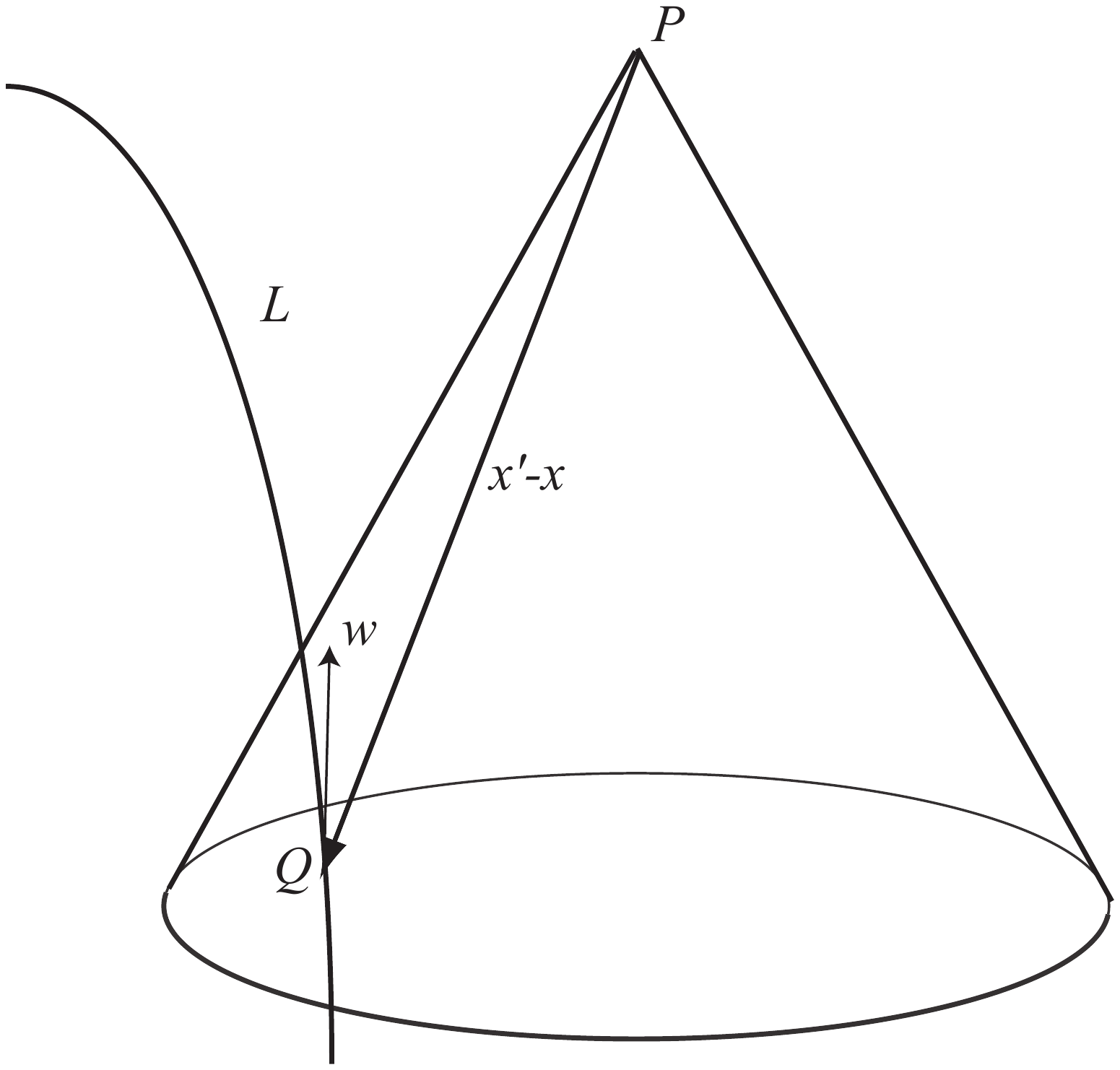}}
%\scalebox{0.4}{\includegraphics{pointChargePoten.eps}}
  \caption{The four-vectors $x'-x,w$ associated with an observer and a moving object.}\label{chargePotent}
\end{figure}

Denote by $P=x^\mu$ the spacetime point in  $K$ at which we want to define the four-potential. Let $L:(x')^\mu (\tau)$ denote the worldline of the object generating our gravitational field. Let the point $Q=x'(\tau(x))$ be the unique point of intersection of the past light cone at $P$ with the worldline $x'(\tau)$, see Figure \ref{chargePotent}.
The time on the worldline of the object corresponding to this intersection is uniquely determined by the point $P$. It is called the \textit{retarded time} and will be denoted by $\tau(x)$.  Note that radiation emitted at $Q=x'(\tau(x))$
at the retarded time will reach $P$ at time $t=(x^0-(x')^0(\tau(x)))/c$.

The four-potential per unit mass at $P$, in an inertial frame, is given by
\begin{equation}\label{4potFin}
  A_\alpha(P)=\frac{-GM w_\alpha}{(x'(\tau(x))-x)\cdot w},
\end{equation}
where $x'(\tau(x))-x$ is the relative position of $P$ and the position of the object at the retarded time and $w$ is its four-velocity at that time. Formula (\ref{4potFin}) certainly holds in an inertial frame in which $M$ is at rest. By Lorentz invariance, this formula holds in any inertial frame. The derivation parallels that of the Li\'{e}nard-Wiechert potentials of the field of an arbitrarily moving charge (see, for example, \cite{LL}, pp. 174-6).

For a gravitational field generated by several objects of masses $M_j$  with worldlines $x'_j(\tau)$, we have to define first their retarded times $\tau_j(x)$, their relative positions $x'_j(\tau_j(x))-x$, and their four-velocities $w_j$ at the retarded time. Then, the total gravitational  four potential at $P$ is
\begin{equation}\label{4potFin2}
  A_\alpha(P)=\Sigma_j \frac{-GM_j(w_j)_\alpha}{(x'_j(\tau_j(x))-x)\cdot w_j}.
\end{equation}
Note that this four-potential tends to $0$ when $P$ is far removed from the sources.

Denote the time and spatial components of the four potential as
\begin{equation}\label{4potdcomp}
   A_\alpha= (\Phi,-\mathbf{A}),
\end{equation}
where the spatial part has a minus sign because it is a co-vector.
Similarly to (\ref{u_def}), we introduce the \textit{dimensionless four-potential}
\begin{equation}\label{a_def}
a_\alpha(P)=\frac{-2A_\alpha (r)}{c^2}=\left(\frac{-2\Phi}{c^2},\frac{2\mathbf{A}}{c^2}  \right)=\left(u,\frac{2\mathbf{A}}{c^2}  \right),
\end{equation}
where $u$ is the dimensionless potential, defined above.

The relativistic force of a field, as in electrodynamics,  acts  via an  antisymmetric  tensor $F_{\alpha\beta}(P)$, generated by the four potential as
\begin{equation}\label{ForceTensor}
  F_{\alpha\beta}(P)=A_{\alpha,\beta}(P)-A_{\beta,\alpha}(P),
\end{equation}
and the $SR$  four-force  of an object  with four-velocity $w$  at $P$ is defined by
\begin{equation}\label{accel}
  F^\alpha(P)=F^\alpha_\beta w^\beta.
\end{equation}
Denote by $\mathbf{E}(P)$ the force acting on an object of unit mass at rest at $P$. Then $\mathbf{E}(P)= F_{\alpha 0}(P)$. As shown in \cite{FSbook}, only the field corresponding to the electric field $\mathbf{E}$ and not the magnetic field
needs to be  corrected in process of SR relativization of electro-dynamics.

Using this idea and following ideas from \cite{Mash}, to define an analog of the metric (\ref{normRND})  for the non-static field. Introduce first a normalized vector
  \begin{equation}\label{n-def2}
 \mathbf{n}(P) = \frac{\mathbf{E}(P)}{|\mathbf{E}(P)|}
\end{equation}
 in the direction of the gravitational force. Let $d\mathbf{x}_n=(d\mathbf{x}\cdot\mathbf{n})\mathbf{n}$ and  $d\mathbf{x}_{tr}=d\mathbf{x}-(d\mathbf{x}\cdot\mathbf{n})\mathbf{n}$,  respectively, denote the projections of the spatial increment $d\mathbf{x}$ in the parallel and transverse directions to $\mathbf{n}(P)$. With this notation, our metric (\ref{normRND}) becomes
 \begin{equation}\label{normRND2}
 ds^2=(1-\frac{2\Phi}{c^2})c^2dt^2+ \frac{2}{c}\mathbf{A}\cdot d\mathbf{x}dt-\frac{1}{1-\frac{2\Phi}{c^2}} d\mathbf{x}_n^2-d\mathbf{x}_{tr}^2.
\end{equation}

This is an  $RND$ metric for a collection of sources.  Since the four-potential tends to $0$ when $P$ is far removed from the sources, this metric is asymptotically flat. One may consider the metric (\ref{normRND2}) as a
 ``correction" of Mashhoon's metric \cite{Mash}, in the same way as the $RND$ metric obtained in \cite{FSS} is a correction of a Newtonian metric. This correction reflects the fact that the potential affects spacetime only
 in the direction of the force and not in the direction transverse to the force. We hope to extend the metric (\ref{normRND2}) to be fully Lorentz covariant.

%The evolution equation under such field is
%\begin{equation}\label{evolMany}
% m\ddot{\mathbf{x}}=-\nabla \Phi-2\frac{\dot{\mathbf{x}}}{c}\times\mathbf{ B}-\nabla \Phi\frac{\dot{\mathbf{x}}^2_{tr}}{c^2}+ 2\frac{\Phi(\mathbf{x})}{c^2}(\dot{\mathbf{x}}\cdot\dot{\mathbf{n}})\mathbf{n},
%\end{equation}
%where $\mathbf{B}=\nabla\times \mathbf{A}$.

\section{Discussion}\label{disc}
$\;$

In this paper we introduced the relativity of spacetime in order  to apply Riemann's principle of ``force equals geometry" to the dynamics under  any static, conservative force. We accomplished this by describing the geometry of the spacetime
of a moving object via a metric derived from the potential of the force field acting on the object. Since an inanimate object has no internal mechanism with which to change its velocity,   it has constant velocity in \emph{its} own world. This led
us to formulate our new \emph{Generalized Principle of Inertia}, which states that: \textbf{An inanimate object moves freely, that is, with zero acceleration, in \emph{its own} spacetime, whose geometry is determined by all of the forces affecting it}.

This is a generalization, or more accurately, a relativization of what Einstein accomplished. In $GR$, an object freely falling in a gravitational field is in free motion. Along a geodesic, the acceleration is zero. The Generalized Principle of Inertia
states that \emph{every} object is in free motion in \emph{its} own world, determined by the forces which affect it. Thus, we assumed the existence of a metric with respect to which the length of the object's trajectory is extremal, enabling us
to use a variational principle and conserved quantities to calculate trajectories.

We constructed the metric of an object moving in a static, spherically symmetric, attractive  force field. We placed an imaginary observer far away from the sources of the field and since he is not affected by the forces, we assumed that he is in an inertial frame $K$ with the Minkowski metric. Using the symmetry of the problem and angular momentum conservation, we were able to identify the part of the metric transverse to the radial direction and show that it is identical to the one in $K$. The remaining part of the metric was obtained from considering radial motion. We defined the metric so that for radial motion, the proper acceleration is independent of the velocity. We used the classical limit to complete the derivation of the metric.  The dynamics built on this metric is called \emph{Relativistic Newtonian Dynamics (RND)}.
%We derived the \textit{dimensionless energy conservation equation} (\ref{ConsRNDgen}) and the \textit{dimensionless equation of motion} (\ref{eomRND}) of $RND$, for both massless particles and objects with non-zero mass.

%It is clear from these equations that this dynamics reduces in the low velocity, weak field limit to classical Newtonian dynamics.

For a gravitational field of a spherically symmetric massive body, our metric is the Schwarzschild metric.  Moreover, $RND$ exactly reproduces the tests of $GR$.
Our theory also handles partially the non-static case.

%%Only second order tests of $GR$ will be able will be able to determine the validity of $RND$.
%The derivation of our metric is much simpler than in $GR$ and uses potentials defined by the sources via Poisson's equation. We expect $RND$ to be useful for studying relativistic gravitational astrophysical (or other) phenomena.

Our methodology here differs from the standard approach in the way we define the radial coordinate $r$. The standard approach (see \cite{Rindler,Rindler2}, for example) is to \emph{define} $r$ so that a sphere of radius $r$ has surface area $4\pi r^2$. Our approach, on the other hand, is to measure $r$ in the inertial frame at infinity. This is a more natural way to define the radial coordinate, since, in practice, this is what is measured.

%\acknowledgments{I wish to thank }

\end{document}